\documentclass[useAMS,usenatbib]{mn2e}

\usepackage{amssymb,amsmath,upgreek,bm,color}

\usepackage{epsfig,graphicx,times,subfigure}
\usepackage[colorlinks=true, citecolor=cyan,
 linkcolor=cyan, menucolor=blue, urlcolor=blue,
 linkbordercolor={0 0 1}, frenchlinks=True, breaklinks]{hyperref}

\newcommand{\kms}{{km~s$^{-1}$}}

\newcommand{\apj}{ApJ}

\newcommand{\oiii}{\hbox{[O\,{\sc iii}]}}
\newcommand{\ha}{\hbox{H$\alpha$}}
\newcommand{\hb}{\hbox{H$\beta$}}
\newcommand{\oii}{\hbox{[O\,{\sc ii}]}}
\newcommand{\sii}{\hbox{[S\,{\sc ii}]}}
\newcommand{\nii}{\hbox{[N\,{\sc ii}]}}

\usepackage{gensymb}

\title[Enhanced Star Formation in Galactic-scale Outflows]
{SDSS-IV MaNGA: Enhanced Star Formation in Galactic-scale Outflows}

\author[Bao, M. et al.]{
\parbox[t]{\textwidth}{\raggedright
Min Bao$^{1,2}$,
Yanmei Chen$^{2}$\thanks{E-mail: chenym@nju.edu.cn},
Qirong Yuan$^{1}$\thanks{E-mail: yuanqirong@njnu.edu.cn},
Yong Shi$^{2}$,
Dmitry Bizyaev$^{3,4}$,
Xiaoling Yu$^{2}$,
Shuai Feng$^{5,6,7}$,
Xiao Cao$^{2}$,
Yulong Gao$^{2}$,
Qiusheng Gu$^{2}$,
Ying Yu$^{2}$
}\\
\vspace*{6pt}\\
$^{1}$School of Physics and Technology, Nanjing Normal University, Nanjing 210023, China\\
$^{2}$Department of Astronomy, Nanjing University, Nanjing 210093, China\\ 
    Key Laboratory of Modern Astronomy and  Astrophysics (Nanjing University), Ministry of Education, Nanjing 210093, China\\
    Collaborative Innovation Center of Modern Astronomy and Space Exploration, Nanjing 210093, China\\
$^{3}$Apache Point Observatory and New Mexico  State University, P.O.  Box 59, Sunspot, NM,  88349-0059,  USA\\
$^{4}$Sternberg Astronomical Institute, Moscow State University, Moscow, Russia\\
$^{5}$Department of Physics, Hebei Normal University, 20 South Erhuan Road, Shijiazhuang,   050024, China\\
$^{6}$Key Laboratory for Research in Galaxies and Cosmology, Shanghai Astronomical Observatory, Chinese Academy of Sciences, \\
80 Nandan Road, Shanghai 200030, China\\
$^{7}$University of the Chinese Academy of Sciences, No.19A Yuquan Road, Beijing 100049, China
	}

\begin{document}

\pagerange{\pageref{firstpage}--\pageref{lastpage}} \pubyear{}

\maketitle
\newpage
\label{firstpage}
\newpage
\pagebreak
\begin{abstract}
Using the integral field unit (IFU) data from Mapping Nearby Galaxies at Apache Point Observatory (MaNGA) survey, we collect a sample of 36 star forming galaxies that host galactic-scale outflows in ionized gas phase. The control sample is matched in the three dimensional parameter space of stellar mass, star formation rate and inclination angle. Concerning the global properties, the outflows host galaxies tend to have smaller size, more asymmetric gas disk, more active star formation in the center and older stellar population than the control galaxies. Comparing the stellar population properties along axes, we conclude that the star formation in the outflows host galaxies can be divided into two branches. One branch evolves following the inside-out formation scenario. The other locating in the galactic center is triggered by gas accretion or galaxy interaction, and further drives the galactic-scale outflows. Besides, the enhanced star formation and metallicity along minor axis of outflows host galaxies uncover the positive feedback and metal entrainment in the galactic-scale outflows. Observational data in different phases with higher spatial resolution are needed to reveal the influence of galactic-scale outflows on the star formation progress in detail.
\end{abstract}

\begin{keywords}
galaxies: star formation - galaxies: abundances - galaxies: kinematics and dynamics

\end{keywords}

\section{Introduction}
\label{introduction}

The regulation of star formation and building-up of stellar mass are hot topics and major unresolved issues in the research of galaxy formation and evolution \citep{2020ARAA..58...99S}. Galactic-scale outflows together with gas accretion play critical role in modulating the star formation activity and interacting with the interstellar medium (ISM) in galaxies \citep{2005ARAA..43..769V}.

More than half a century ago, \cite{1963ApJ...137.1005L} demonstrated the existance of explosion in the galactic center of M82. Many physical mechanisms can invoke the powerful explosion, e.g., stellar winds and supernovae \citep{1990ApJS...74..833H, 2021MNRAS.500.3802H}, as well as active galactic nuclei \citep{2012ARAA..50..455F, 2014AA...562A..21C, 2019MNRAS.490.3830B}. The radiation pressure from photoionization or shock excitation blows gas out along the polar axes \citep{2010ApJ...721..505R, 2014AA...567A.125G, 2017ApJ...834...30F}. More and more studies indicate the ubiquity of the multi-phase galactic-scale outflows at all cosmic epochs \citep{2019ApJ...871...37H, 2019ApJ...873..122D, 2020AA...633A..90G, 2020ApJ...901..151S, 2020ApJ...905..166L, 2021MNRAS.500.3802H}.

However, agreement still didn't reach on the effect of galactic-scale outflows on star formation acticities. On the one hand, the energy injection from the outflows could prevent gas from collapse, which thus restrains star formation and is called negative feedback \citep{2010AA...518L..41F, 2011ApJ...733L..16S, 2012MNRAS.425L..66M}. On the other hand, the outward winds extrude the gas clouds on the path and improve the volume density of molecular gas \citep{2012AA...537A..44A, 2015AA...574A..85A, 2009ApJ...700L.104S}. Such a procedure could accelerate the formation of stars and is called positive feedback. As a matter of fact, these two schemes could coexist in one case, but correspond to different local environment \citep{2019ApJ...881..147S, 2021AA...645A..21G}.

There exist abundant observational evidences of the positive feedback on star formation in host galaxies, while few proofs demonstrate the enhanced star formation in the galactic-scale outflows. \cite{2017Natur.544..202M} reported spectroscopic observations that reveal star formation occurring in a galactic outflow at a redshift of 0.0448. \cite{2019MNRAS.485.3409G} pointed out that star formation occurs inside at least 30 percent of the galactic-scale outflows sample in the MaNGA survey. However, statistically spatial resolved evidence of the star formation enhancement in the galactic-scale outflows is still missing.

In this work, we construct a sample of star forming galaxies (SFGs) from the MaNGA survey and further select galactic-scale outflows therein. By comparison with the galaxies that don't show outflow features, we summarize the global features of the outflows host galaxies. The common structure of the galactic-scale outflows can be modelled as the famous M82 \citep{1988Natur.334...43B}, where the outward gas looks like bipolar winds and is projected to the minor axis along the line-of-sight. By comparing the stellar population properties in galactic-scale outflows (minor axis) and along stellar disk (major axis), we find robust evidence on positive feedback and metal entrainment in the galactic-scale outflows.

This paper is organized as follows. The sample selection and methodology of data processing are presented in Section \ref{sec: sample}. We describe the influence of the outflows on the host galaxies in Section \ref{sec: influence of outflows on hosts}, which includes both global properties and properties along the axes. Our understandings on these results are discussed in Section \ref{sec: discussion}. Finally, a summary is given in Section \ref{sec: summary}. Throughout this paper, we adopt a set of cosmological parameters as follows: $H_0=70\,{\rm km~s}^{-1}\,{\rm Mpc}^{-1}$, $\Omega_m=0.30$, $\Omega_{\Lambda}=0.70$.

\section{The Sample}
\label{sec: sample}

\subsection{Data and Methodology}
\subsubsection{The Data}
MaNGA is one of three core programs in the fourth-generation Sloan Digital Sky Survey (SDSS-IV) \citep{2015ApJ...798....7B, 2016AJ....152...83L}, it employs the Baryon Oscillation Spectroscopic Survey (BOSS) spectrographs \citep{2013AJ....146...32S} on the 2.5m Sloan Foundation Telescope \citep{2006AJ....131.2332G}. This survey aims to conduct IFU observation for $\sim$ 10,000 nearby galaxies with a flat stellar mass distribution in 10$^{9}-$10$^{11}\rm M_{\odot}$ and redshift $0.01 < z < 0.15$ \citep{2017AJ....154...28B}. Two dual-channel BOSS spectrographs \citep{2013AJ....146...32S} provide simultaneous wavelength coverage from 3600 to 10000\rm\AA. The spectral resolution is $\sim$2000, allowing measurements of all strong emission line species from {\oii}$\lambda$3727 to {\sii}$\lambda$6731. The MaNGA sample and data products used here are drawn from the SDSS Data Release 15 (DR15) \citep{2019ApJS..240...23A}, which includes $\sim$4621 galaxies observed in the first three years of the survey. 

The global stellar mass ($M_{\star}$) and star formation rate (SFR) in this work are extracted from the GALEX-SDSS-WISE Legacy Catalog (GSWLC) and derived through UV/optical SED fitting \citep{2007ApJS..173..267S}. We stack the spectra inside the MaNGA bundle and measure the strength of the 4000\r{A} break ($Dn4000$) to get an idea of the global stellar population age. $Dn4000$ is defined as the ratio of flux density between two narrow continuum bands 3850-3950\r{A} and 4000-4100\r{A}. The bulge-to-total light ratio (fracDeV) is obtained by Sersic+Exponential fits to the $r$-band 2D surface brightness profiles from the MaNGA PyMorph DR15 photometric catalog \footnote{https://www.sdss.org/dr15/data\_access/value-added-catalogs}. The other intrinsic properties, e.g., redshift($z$), effective radius ($Re$), axis ratio ($q=b/a$) and photometric position angle ($\phi$) are adopted from NASA-Sloan Atlas (NAS, \cite{2011AJ....142...31B}).

The spatially resolved spectral properties are obtained from the MaNGA data analysis pipeline (DAP, \cite{2019AJ....158..231W}), e.g., emission line flux, stellar/gas velocity and velocity dispersion, as well as $Dn4000$.  We limit the signal-to-noise ratios of the related emission lines to be higher than 2 for further calculations. The asymmetric feature is the indicator of gas disk stability and we cite the kinematic asymmetry from \cite{2020ApJ...892L..20F}, which use the KINEMETRY package \citep{2006MNRAS.366..787K} to fit the velocity field of the ionized gas. Similarly, we use KINEMETRY package to fit the kinematic position angle of the circular gas disk. The kinematic properties of gas are traced by ionized Hydrogen (i.e., H$\alpha$), while all the emission line centers are tied together in MaNGA DAP. Besides, we extract the mass/luminosity weighted stellar population age and stellar mass in each spaxel from the data cube of P\scriptsize IPE\normalsize 3D \citep{2016RMxAA..52...21S}, which is a fitting tool for the analysis of the spectroscopic properties of the stellar population.

\subsubsection{The Methodology}
Using {\ha}/{\hb} ratio, we estimate extinction corrected H$\alpha$ luminosity under the Case B \citep{2001PASP..113.1449C}:
\begin{equation}
F_{\lambda}=F_{\lambda,0}\times10^{-0.4k(\lambda)E(B-V)},
\label{eq1}
\end{equation}  
where $k(\lambda)$ is the Galactic dust attenuation curve, and color excess $E(B-V)=0.934\times \ln[(F_{H\alpha}/F_{H\beta})/2.86]$.

The SFR in the star forming region is estimated from the {\ha} luminosity as suggested by \cite{1998ARAA..36..189K} with a Salpeter IMF \citep{1955ApJ...121..161S}:
\begin{equation}
SFR(M_{\odot}yr^{-1}) = 7.9\times10^{-42}L_{H\alpha}.
\label{eq2}
\end{equation}    
The spatially resolved star formation rate surface density ($\Sigma_{\rm SFR}$) is estimated by further dividing the physically spaxel area. While the $\Sigma_{\rm SFR, 1kpc}$ is defined as the SFR surface density within the central 1kpc radius.

The gas-phase metallicity is an important indicator for constraining the past star formation history (SFH) of galaxies. Several calibrations, either empirical, theoretical or hybrid, have been proposed to derive gas metallicity from emission line fluxes. We apply the analytical fit from \cite{2004ApJ...613..898T} and estimate gas-phase metallicity using $R_{\rm 23}$ as:
\begin{equation}
12+\log(O/H) = 9.185-0.313x-0.264x^{2}-0.321x^{3},
\label{eq3}
\end{equation}
where $x \equiv \log({\oii}\lambda \lambda3726,3729+{\oiii}\lambda \lambda4959,5007)/{\hb}$. The metallicity within the central 1kpc is denfined as the median of all the spaxels within central 1kpc radius. Other metallicity tracers are also adopted to check the reliability of the metallicity measurements, e.g., ({\oiii}$\lambda$5007/{\hb})$\times$({\ha}/{\nii}$\lambda$6583), {\nii}$\lambda$6583/{\ha} \citep{2013AA...559A.114M}. Those measurements are 0.4$dex$ lower than the metallicity that calculated as equation \ref{eq3}, but the overall trends are consistent. The discrepancy is reasonable considering that equation \ref{eq3} is derived by the modeled calibrators, while \cite{2013AA...559A.114M} calibrated the metallicity with direct method. In general, the former is higher than the later \citep{2017MNRAS.469.2121S, 2017ApJ...844...80B}.

\subsection{Sample Selection}
In this work, we focus on the star formation driven galactic-scale outflows, and the parent sample is selected as follows:
\begin{itemize}
\item[1.] It locates in star forming main sequence (grey hollow circles in Figure \ref{fig1}a). The black solid line, which expressed as $\log$(SFR) = 0.86$\log$($M_{\star}$) - 9.29, is adopted to seperate the star forming main sequence from the other galaxies.
\item[2.] $\log$({\oiii}/{\hb}) $<$ 0.61/($\log$({\nii}/{\ha}) - 0.05) + 1.3. The central spaxel of each galaxy is limited to locate in the star-forming region based on the demarcation proposed by \cite{2003MNRAS.346.1055K}.
\item[3.] $r$-band fracDeV $<$ 0.8. We follow \cite{2008MNRAS.388.1321P} to distinguish disk galaxies (fracDeV $<$ 0.8) from the early-type galaxies (fracDeV $>$ 0.8), so that the estimation of the galaxy inclination angle ($i$) will be feasible.
\item[4.] $i > 45^{\degree}$. The relation between inclination angle and axial ratio is given by $\cos^{2}i = (q^{2} - q_{\rm 0}^{2})/(1 - q_{\rm 0}^{2})$, where $q$ is the axial ratio and $q_{\rm 0}$ = 0.2 \citep{1926ApJ....64..321H}.
\item[5.] The median {\oiii}$\lambda$5007 equivalent width ({\oiii} EQW) of spaxels along the minor axis is higher than that along the major axis.
\end{itemize}
Finally, we select 386 SFGs as the parent sample. We inspect the EQW maps of {\oiii} and {\ha}, and find 36 of them (red stars in Figure \ref{fig1}a) having galactic-scale outflow features in both maps.

Figure \ref{fig2} shows an example of the star-forming galaxies with galactic-scale outflows. Figure \ref{fig2}a is the $g, r, i$ color image from the SDSS survey. Figure \ref{fig2}b and \ref{fig2}c show the BPT diagram and resolved BPT map. In Figure \ref{fig2}b, the black solid curve shows the demarcation between starburst galaxies and AGNs defined by \citet{2003MNRAS.346.1055K} and the black dashed curve shows the demarcation proposed by \citet{2001ApJ...556..121K}. In both Figrue \ref{fig2}b and \ref{fig2}c, the blue pixels mark the star forming region, the green pixels mark the composite region with contributions from both AGN/shock and star formation, the red pixels mark the AGN/shock region. The color scheme is quantified by the distance of each pixel to the dashed curve \citep{2003MNRAS.346.1055K} in Figure \ref{fig2}b. It is clear that this galaxy is star forming dominated. The bipolar structure of outflows is obvious in the {\ha} EQW and {\oiii} EQW maps in Figure \ref{fig2}d and \ref{fig2}e.

The stellar velocity map in Figure \ref{fig2}f shows two counter-rotating stellar disks. The gas velocity map traced by {\ha} in Figure \ref{fig2}g follows a regularly rotated gas disk with multiple velocity components in the bipolar region. Figure \ref{fig2}h shows the best fit of the observed velocity field, which follows the method of \cite{2006MNRAS.366..787K} and assumes circular orbits in a thin disk. We refer the reader to \cite{2008MNRAS.390...93K, 2011MNRAS.414.2923K} for more details. The model is expressed as $V(R,\psi)=V_{\rm C}(R)\ \sin(i)\ \cos(\psi)$, where $R$ is the radius of a circular orbit in the galactic plane, $V_{\rm C}$ is the {\ha} circular velocity at radius $R$ and is defined as the velocity along the major axis in each 0.5arcsec bin, $i$ is the inclination angle and $\psi$ is the azimuthal angle measured from the major axis in the galactic plane with a coverage of [0, 2$\pi$]. We subtract the disk model from {\ha} velocity field and the residual is shown in Figure \ref{fig2}i. The bipolar galactic-scale outflows in Figure \ref{fig2}i is quite obvious. The outward ionized gas moves toward the Earth in the northwest, while moves away in the southeast. The maximum projected velocity of galactic outflows equals 109{\kms}. After inclination angle ($i_{\rm disk}=55^{\degree}$) correction, the actual velocity is larger than 150{\kms}, which means the star formation driven outflows are quite powerful in this case.

\subsection{Control Sample}
\begin{figure*}
 \resizebox{0.9\textwidth}{!}{\includegraphics{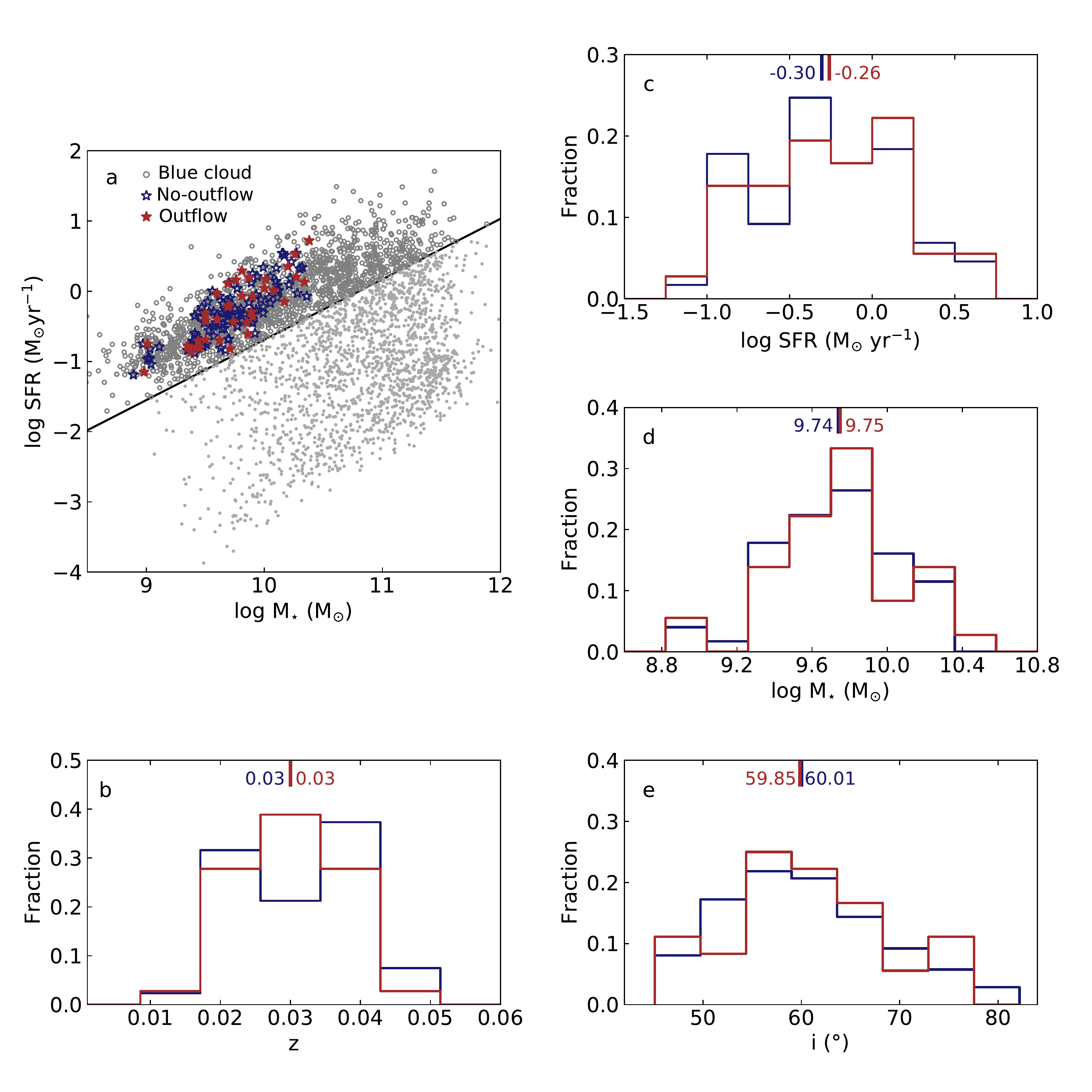}}
\caption{a: Main sequence relation. The black solid line is adopted to seperate the star forming main sequence from the other galaxies. The grey hollow and solid points represent the star forming (SF) galaxies and non-SF galaxies. The red and blue stars mark the sample and control galaxies. b: Redshift distributions. The red and blue histograms represent the distributions of sample and control galaxies. The red and blue bars mark the medians of the corresponding distributions. Without supplementary description, the same color coding is hereinafter used in the following figures. c: Star formation rate distributions. d: Stellar mass distributions. e: Inclination angle distributions.}
\label{fig1}
\end{figure*}

\begin{figure*}
 \resizebox{0.99\textwidth}{!}{\includegraphics{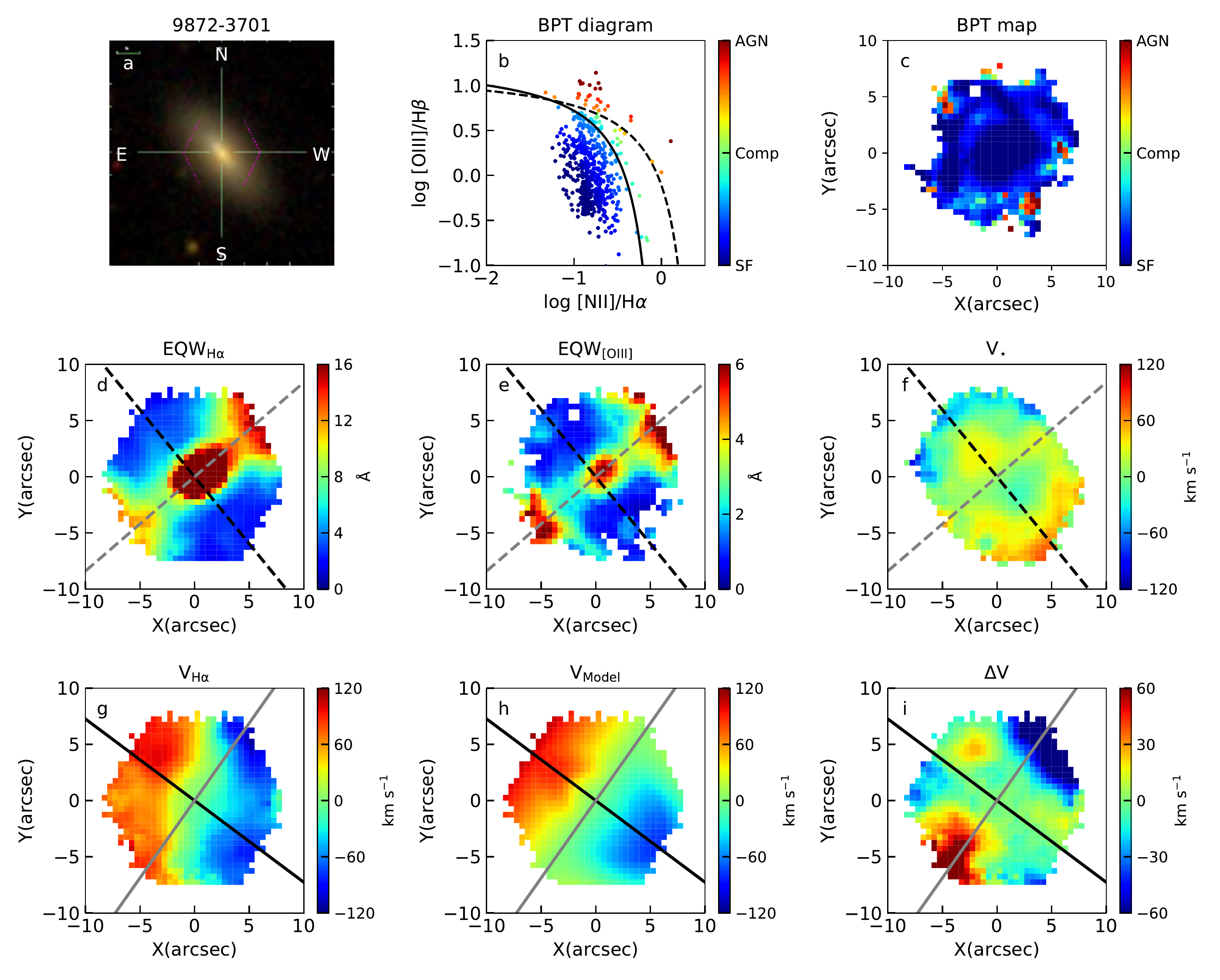}}
\caption{a: SDSS $g, r, i$ colour image. b: BPT diagram. The black solid curve shows the demarcation between starburst galaxies and AGNs defined by \citet{2003MNRAS.346.1055K}, the black dashed curve shows the demarcation proposed by \citet{2001ApJ...556..121K}. The blue pixels designate the star forming region, the green pixels mark the composite region with contributions from both AGN/shock and star formation, the red pixels represent the AGN/shock region. c: The spatially resolved BPT diagram. The color definition is same as panel b. d: Equivalent width map of {\ha}, the black dashed line marks the photometric major axis, while the grey dashed line marks the photometric minor axis; e: Equivalent width map of {\oiii}$\lambda5007$, the black and grey dashed lines are same as panel d. f: Velocity map of stars, the black and grey dashed lines are same as panel d. g: Velocity map of {\ha}, the black solid line marks the kinamatic major axis, while the grey solid line marks the kinematic minor axis. h: Circular disk model, the black and grey solid lines are same as panel g. i: Residual ($V_{\rm H\alpha} - V_{\rm Model}$) after subtracting the disk model from the H$\alpha$ velocity field, the black and grey solid lines are same as panel g.}
\label{fig2}
\end{figure*}

In order to quantify the influence of outflows on the evolution of host galaxies, we build a control sample of galaxies without outflow features. For each star-forming galaxy with galactic-scale outflows (SFO), we select 5 control galaxies (CGs, blue stars in Figure \ref{fig1}a) which are closely matchd in three dimensional space of stellar mass ($|\Delta \log M_{\star}| < 0.1$), star formation rate ($|\Delta \log \rm SFR| < 0.2$) and inclination angle ($|\Delta i| < 5 \degree$). The distributions of those properties of SFOs and CGs are displayed in Figure \ref{fig1}c, \ref{fig1}d and \ref{fig1}e. Moreover, the redshift distributions of SFOs and CGs (see Figure \ref{fig1}b) are consistent.

The motivation for choosing these three matching parameters is the following: (i) stellar mass is the most fundamental parameter and many other physical properties are known to vary strongly with stellar mass; (ii) matched SFR implies similar activity in SFOs and CGs to probe the driven mechanism of outflows; (iii) the inclination angle constraint ensures similar projection effects on SFOs and CGs.

\section{INFLUENCE OF OUTFLOWS ON HOSTS}
\label{sec: influence of outflows on hosts}

\subsection{Global Properties}
The comparison in the global properties between SFOs and CGs can help explore the origin of galactic-scale outflows. The parameters concerned in this section include: galaxy size traced by effective radius, kinematic stability traced by asymmetry of gas velocity map and stellar population properties traced by star formation rate surface density or stellar population age.

\begin{figure*}
 \resizebox{0.8\textwidth}{!}{\includegraphics{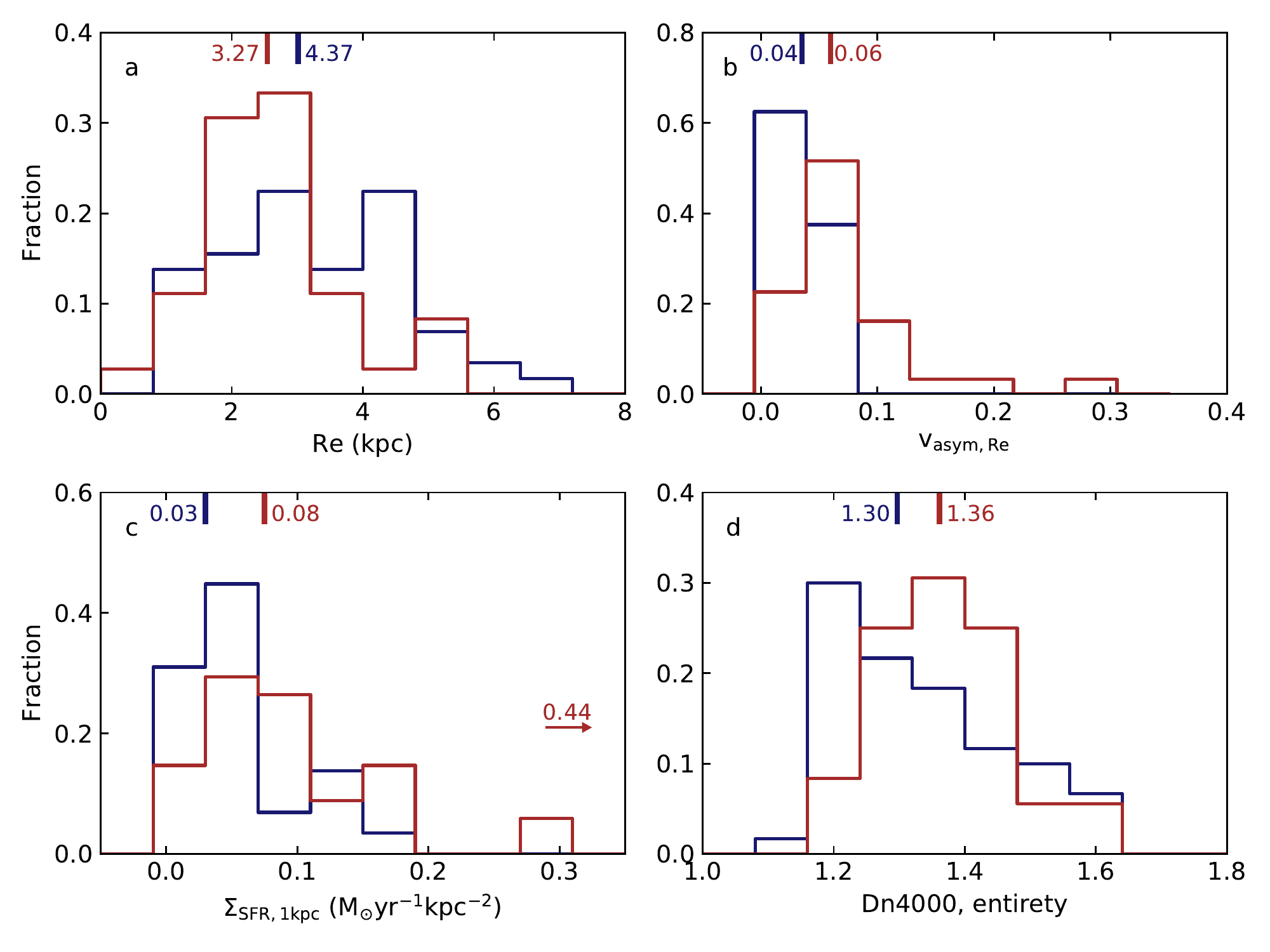}}
\caption{a: Effective radius distributions. b: Gas velocity field asymmetry distributions. c: Central (1kpc) star formation rate surface density distributions. d: Lighted weighted stellar population age index ($Dn4000$) distributions.}
\label{fig3}
\end{figure*}

In Figure \ref{fig3}, the red/blue histogram represents the distribution of physical properties of SFOs/CGs, while the red/blue bar on the top of each panel shows the median of each parameter of SFOs/CGs. Figure \ref{fig3}a shows that the effective radius of SFOs is smaller than the CGs. We obtain both $R_{\rm 50}$ (Petrosian 50\% light radius) and $R_{\rm 90}$ (Petrosian 90\% light radius) from the NYU value-added catalog \citep{2005AJ....129.2562B} and calculate the concentration ($C = R_{\rm 90}/R_{\rm 50}$). The distributions of $C_{\rm SFO}$ and $C_{\rm CG}$ match each other, thus SFOs tend to be smaller in physical sizes compared to the CGs.

The asymmetry of gas velocity field can indicate the turbulance in gas disk. Previous studies have reported that properties of ionized gas at the effective radius are representative of the average properties of galaxies \citep{2015AA...581A.103G, 2016RMxAA..52..171S}, thus we choose the asymmetry at effective radius ($v_{\rm asym,Re}$) for comparison. The distributions in figure \ref{fig3}b show higher asymmetric gas kinemtics in the SFOs, which could be contributed by gas accretion or galaxy interaction.

Observationally, galactic-scale outflows are a general consequence of high SFR surface density \citep{2002ApJ...577..691H, 2010AJ....140..445C}. In Figure \ref{fig3}c, we compare the $\Sigma_{\rm SFR, 1kpc}$ between SFOs and CGs and find that the central regions of SFOs are more active than the CGs. The active star formation implies the birth of young stellar population. However, the global 4000\rm\AA\ break in SFOs turns out to be higher than CGs (Figure \ref{fig3}d), which indicates older global stellar population.

In terms of the galactic-scale outflows host galaxies, the higher global $Dn4000$ implies the generally older stellar population, while the higher $\Sigma_{\rm SFR, 1kpc}$ invokes ongoing star formation in the galactic center. Considering that SFOs have similar SFR but smaller physical size than CGs, they could go through strong star formation recently. The centrally concentrated star formation in SFOs can be triggered by gas accretion or galaxy interaction, which is shown as asymmetric gas disk. Besides, the radiation pressure from the star formation further drives the powerful galactic-scale outflows. 

\subsection{Enhanced Star Formation and Metallicity in Outflows} 
\label{sec: enhanced star formation and metallicity in outflows}

In this section, we focus on comparing gas-phase metallicity, star formation activity, as well as stellar population age in galactic-scale outflows (minor axis) and along stellar disk (major axis). Through this kind of comparisons, we would have an idea about how the star formation driven galactic-scale outflows influnce the evolution of host galaxies.

\begin{figure*}
 \resizebox{0.63\textwidth}{!}{\includegraphics{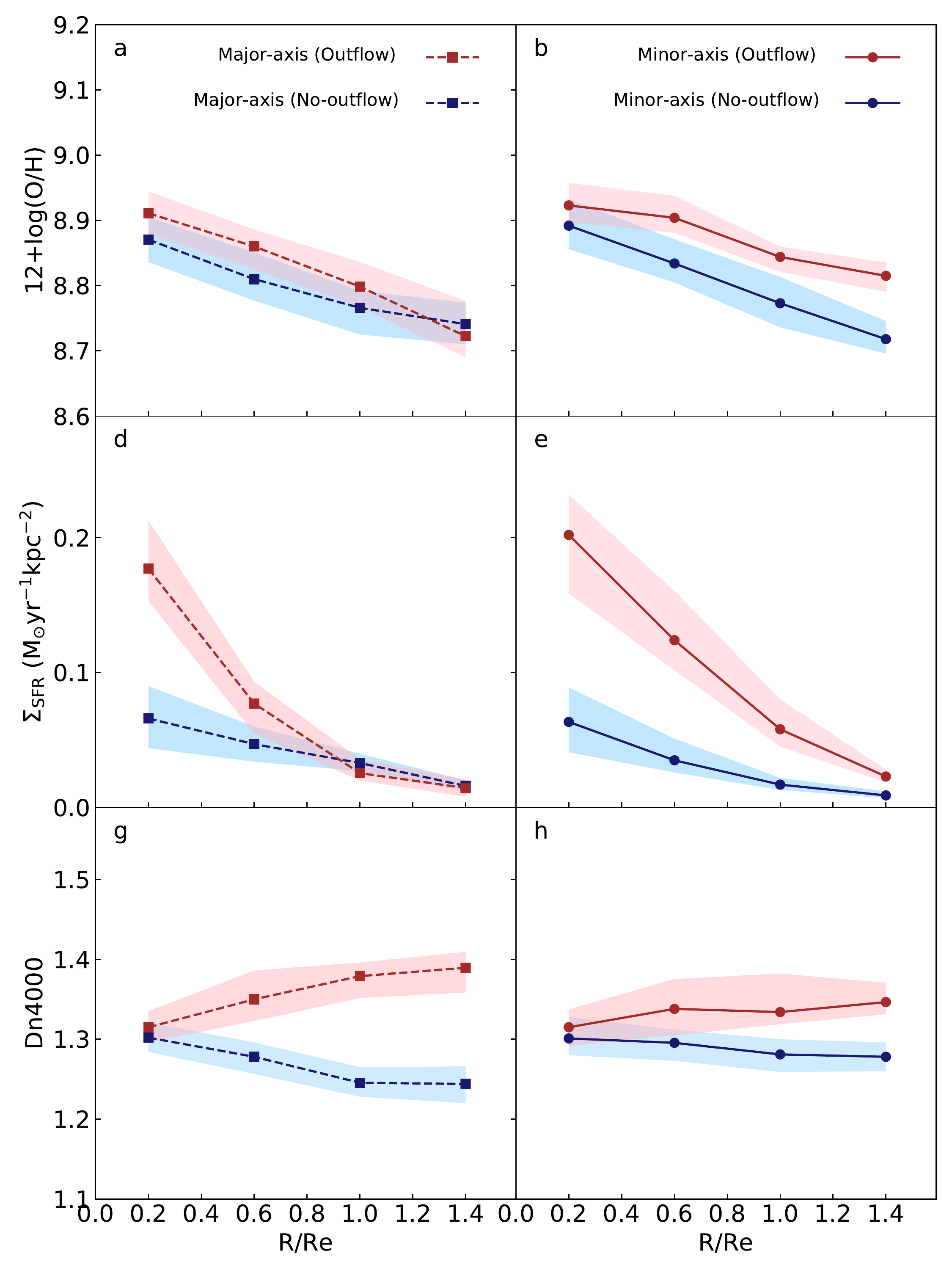}}
 \resizebox{0.35\textwidth}{!}{\includegraphics{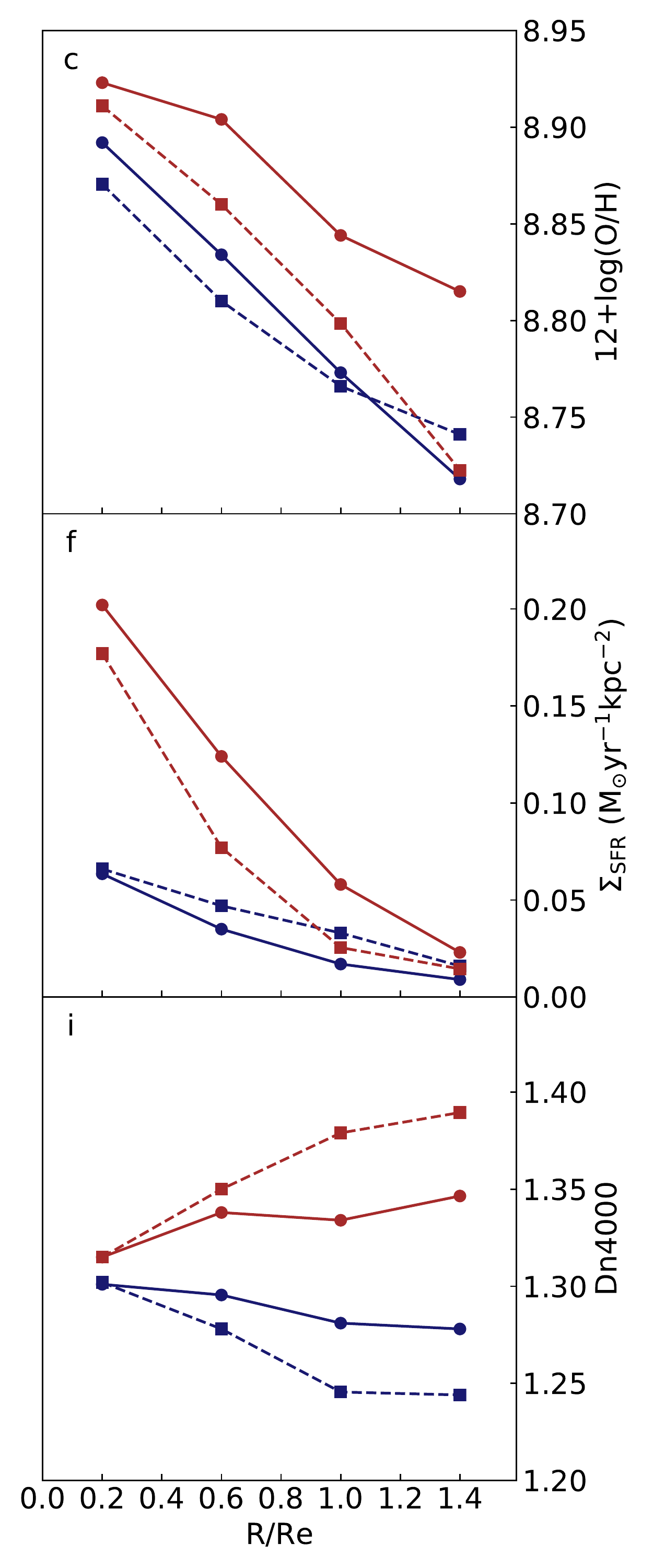}}
\caption{The red dashed/solid profile represents the gradient along the major/minor axis of SFOs, on which the red squares/circles mark the median in each 0.4$Re$ bin. The pink shadows covers 40\%-60\% error bar range around the medians in SFOs. The blue dashed/solid profile represents the gradient along the major/minor axis of CGs, on which the blue squares/circles mark the median in each 0.4$Re$ bin. The sky-blue shadows covers 40\%-60\% error bar range around the medians in CGs. The last panel in each row gathers the gradients in SFOs and CGs. a b c: The gas phase metallicity gradients. d e f: The star formation rate surface density gradients. g h i: The light weighted stellar population age index ($Dn4000$) gradients.}
\label{fig4}
\end{figure*}

From top to bottom, Figure \ref{fig4} shows the gradients of gas-phase metallicity, star formation rate surface density and indicator of light weighted stellar population age ($Dn4000$). The galactocentric distance is normalized to the effective radius and binned in units of 0.4$Re$. In the left column of Figure \ref{fig4}, the red/blue square marks the median in each radial bin along the major axis of SFOs/CGs. The red/blue dashed profile represents the radial gradient along the major axis of SFOs/CGs. In the middle column of Figure \ref{fig4}, the red/blue circle marks the median in each radial bin along the minor axis of SFOs/CGs. The red/blue solid profile represents the radial gradient along the minor axis of SFOs/CGs. The pink/sky-blue shadow in those two columns covers 40\%-60\% error bar range around the gradient of SFOs/CGs. Besides, the profiles of each property are collected in the corresponding panel in the right column of Figure \ref{fig4}.

The gas-phase metallicity along major axis of SFOs is higher than CGs in Figure \ref{fig4}a, and the gradients are similar. The metallicity along minor axis of SFOs is also higher than CGs in Figure \ref{fig4}b, while the gradient in SFOs is shallower. In Figure \ref{fig4}c, the slopes of the gradients along major and minor axes of CGs, as well as gradient along major axis of SFOs approximately equal 0.1$dex/Re$, which is coincident with the typical metallicity gradient of star forming galaxies \citep{2020ApJ...890L...3S}. The metallicity in the central 0.4$Re$ of both SFOs and CGs are approximative. The gradients along major and minor axes of CGs are consistent in each radial bin. However, in the outer 0.4$Re$ of SFOs, the metallicity along minor axis is 0.15$dex$ higher than major axis. Those enriched metals along minor axis of SFOs could be entrained by the galactic-scale outflows.

Star formation rate describes the activity of ongoing star formation which happens in the recent $10^{6-7}yr$. The star formation rate surface density along major axis of SFOs is higher than CGs (Figure \ref{fig4}d). Considering the similar SFR in SFOs and CGs (Figure \ref{fig1}b), the smaller physical size of SFOs (Figure \ref{fig3}a) causes denser star formation activities. Besides, the $\Sigma_{\rm SFR}$ in central 0.4$Re$ of SFOs is obviously higher than the outer 0.4$Re$, while such difference is small in CGs. This tendency further proves that the active star formation in the center of SFOs could be the trigger mechanism for the powerful outflows. Similar gradients along minor axes are shown in Figure \ref{fig4}e. In Figure \ref{fig4}f, the $\Sigma_{\rm SFR}$ gradients along major and minor axes of CGs are consistent. While the $\Sigma_{\rm SFR}$ along minor axes of SFOs is entirely higher than major axis, which is a direct evidence for the star formation enhancement in the galactic-scale outflows.

Light weighted stellar population age, which can be indicated by $Dn4000$, traces the generation of young stellar population in galaxies. Such a parameter implies the average intensity of star formation in $10^{9}yr$ time scale. In Figure \ref{fig4}g and \ref{fig4}h, $Dn4000$ along both major and minor axes of SFOs are higher than CGs, which means that the global stellar population in SFOs is older. On this basis, the similar global SFR in SFOs and CGs implies a recent star formation in SFOs that may be triggered by gas accretion or galaxy interaction. Comparing the $Dn4000$ gradients in Figure \ref{fig4}i, we find that the stellar population along the minor axis of SFOs is younger than the major axis. The existance of younger stellar population along the minor axis of SFOs further supports the enhanced ongoing star formation in the galactic-scale outflows.

\section{DISCUSSION}
\label{sec: discussion}

A plenty of physical processes impact the star formation activities in the galaxy evolution, e.g., gas inflows, gas outflows and galaxy interaction. To better understand this complex system, one of the key clues is to study the stellar population content at the current stage \citep{2010AJ....140..445C, 2019ApJ...882..145B}.

IFU data have been wildly used to analyse the radial gradients of the stellar population parameters in the spatially resolved manners \citep{2015AA...581A.103G, 2019MNRAS.484.5009E, 2020ApJ...896...75S}. In this work, we collect SFGs that host galactic-scale outflows from the MaNGA survey. The active star formation but older stellar population along axes of SFOs suggest that the formation of stars in SFOs can be classified into two branches. One branch lodges in stellar disk and steadily forms following an inside-out process. The other dominates over the galactic center and could be triggered by gas accretion or galaxy interaction recently.

\subsection{Inside-out Formation} 
\label{sec: inside-out formation}

The scenario that pristine gas is accreted from cosmic web or satellites \citep{2006ApJ...645..986R, 2007MNRAS.374.1479G} and primarily collapses in the galactic center, is so called inside-out formation. In such a process, stars in the galactic center are formed prior to that in the outer disk \citep{2020AJ....159..195D}, thus the stellar population in the central region is expected to be older \citep{2020MNRAS.495.3387P}.

\begin{figure*}
    \resizebox{0.99\textwidth}{!}{\includegraphics{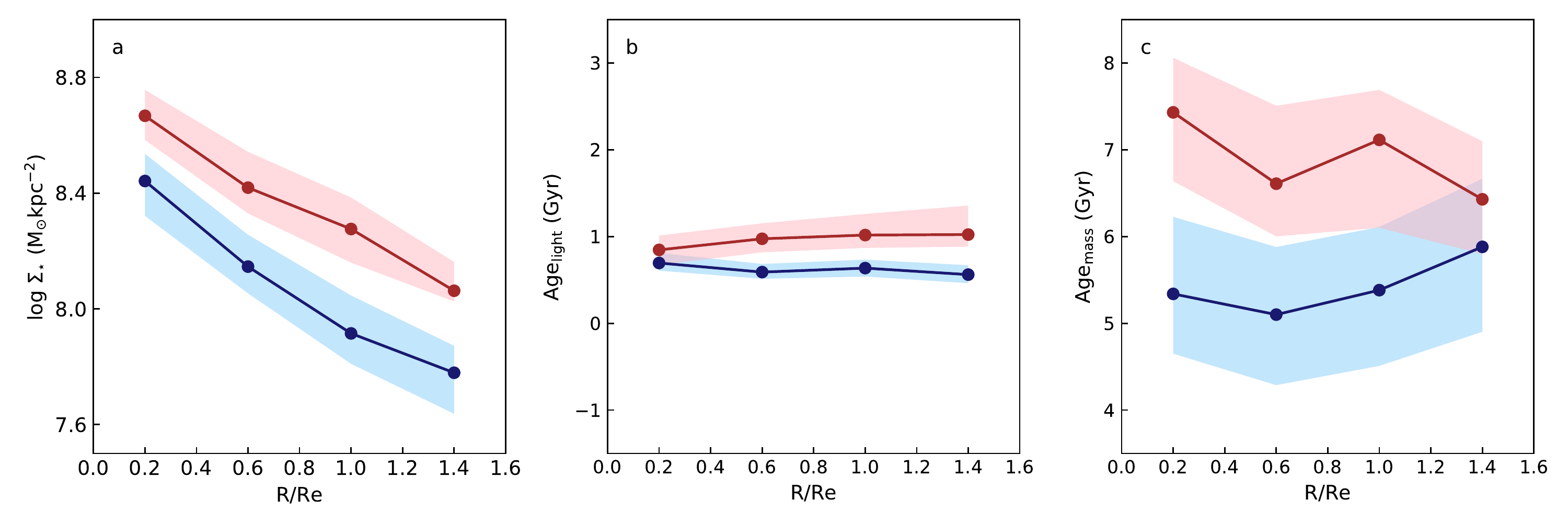}}
   \caption{The color and mark settings are same as Figure \ref{fig4}. a: The stellar mass surface density gradients along the minor axis. b: The light weighted age gradients along the minor axis. c: The mass weighted age gradients along the minor axis.}
   \label{fig5}
   \end{figure*}

The negative $Dn4000$ gradients along axes of CGs (blue profiles in Figure \ref{fig4}i) follow the inside-out formation, while such gradients reverse in SFOs (red profiles in Figure \ref{fig4}i). Some studies suggested evidence for an outside-in formation process, but only in low-mass dwarf galaxies \citep{2015ApJ...804L..42P, 2018ApJ...857...63S}. As shown in Figure \ref{fig5}a, the negative $\Sigma_{\star}$ gradients imply that stars formed earlier or more efficient in the central region of both SFOs and CGs. Bisides, the negative metallicity gradients in SFOs and CGs (Figure \ref{fig4}c) could be explained as the metal-rich stars assembling in the galactic center and metals being released to ISM by stellar winds or supernovae. Those stellar population property gradients are consistent with the inside-out formation scenario.

Figure \ref{fig5}b and \ref{fig5}c show the light and mass weighted stellar population age gradients, which trace the young and old stellar populations, respectively. Both the light and mass weighted stellar population ages in SFOs (red profiles in Figure \ref{fig5}b and \ref{fig5}c) are older than CGs (blue profiles in Figure \ref{fig5}b and \ref{fig5}c), while the difference in mass weighted stellar population age is more prominent. Therefore, the old stellar population dominates the stellar components in SFOs.

In Figure \ref{fig4}f, the $\Sigma_{\rm SFR}$ along axes of SFOs (red profiles) is overall higher than CGs (blue profiles), which implies denser star formation in the outflows host galaxies recently. Besides, the steeply negative $\Sigma_{\rm SFR}$ gradients in SFOs suggest that the strong star formation is focus on the galactic center. As shown in Figure \ref{fig4}i, it is the newborn stars that pull down the light weighted stellar population age in the center of SFOs.

\subsection{Recent Star Formation}
\label{sec: recent star formation}

\begin{figure*}
    \resizebox{0.99\textwidth}{!}{\includegraphics{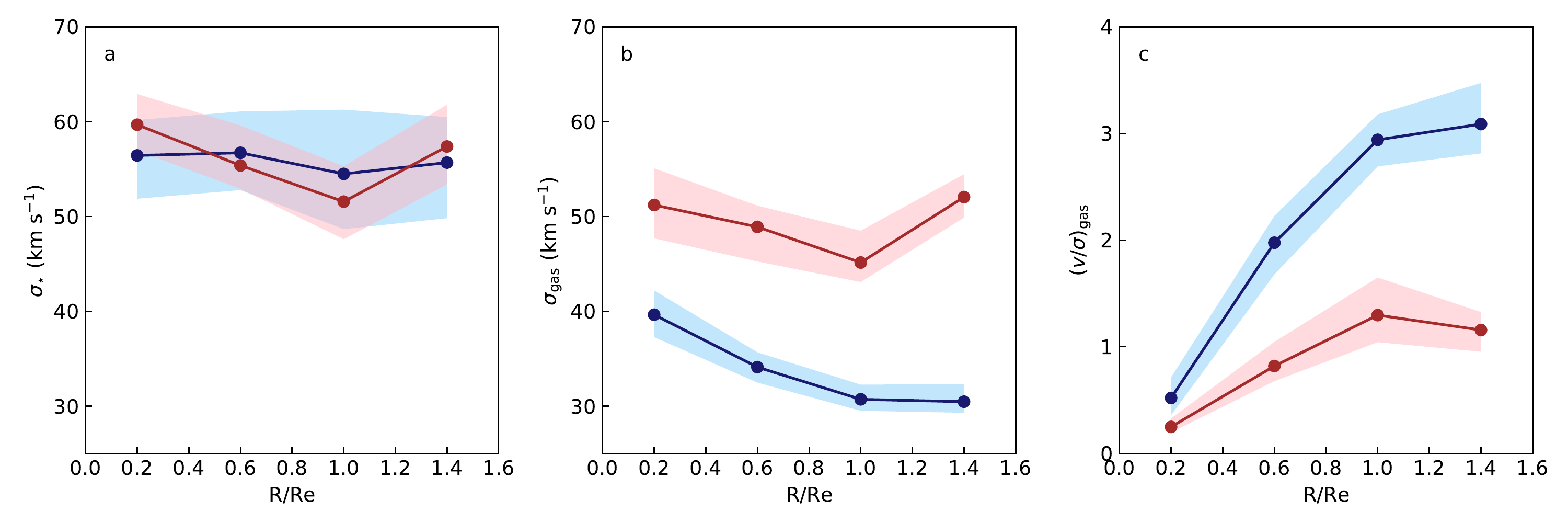}}
   \caption{ The color and mark settings are same as Figure \ref{fig4}. a: The stellar velocity dispersion gradients along the minor axis. b: The gas velocity dispersion (traced by {\ha}) gradients along the minor axis. c: The gas velocity to velocity dispersion ratio (traced by {\ha}) gradients along the minor axis.}
\label{fig6}
\end{figure*}

Figure \ref{fig6}a shows the similar stellar velocity dispersion ($\sigma_{\star}$) along minor axis of SFOs and CGs, while Figure \ref{fig6}b shows higher gas velocity dispersion ($\sigma_{\rm gas}$), traced by ionized Hydrogen (H$_{\alpha}$), along the minor axis of SFOs. The enhanced turbulance in the ionized gas of SFOs, could be a conbined result of gas accretion, gas outflows and galaxy interaction. Such turbulance, which also behaves as more asymmetric gas kinematics in SFOs in Figure \ref{fig3}b, is a key factor in triggering the recent star formation.

We calculate the fraction of kinematically misaligned galaxies in SFOs and CGs, which are 22\% and 7\%. The fraction in CGs is consistent with the typical fraction in MaNGA survey \citep{2016MNRAS.463..913J}, while the fraction in SFOs is three times higher. Simulations show that continuous gas accretion \citep{1998ApJ...506...93T} or galaxy interaction \citep{2014MNRAS.444.3357N} can inject kinematically decoupled gas. In Figure \ref{fig6}c, the gas velocity to velocity dispersion ratio $(v/\sigma)_{\rm gas}$, traced by ionized Hydrogen (H$\alpha$), in SFOs (red profile) is lower than CGs (blue profile). This implies the consumption of gas angular momentum in SFOs, which is the result of the collision between the pre-existing and accreted gas. Under gravitation, the slow-rotating gas in SFOs would be accreted into the central region and trigger star formation.

The star formation in the center of SFOs produces strong radiation pressure and further drives powerful outflows. In Figure \ref{fig4}f, the $\Sigma_{\rm SFR}$ along the minor axis (red solid profile) of SFOs is higher than the major axis (red dashed profile), which demonstrates star formation enhancement in the galactic-scale outflows. This enhancement, known as positive feedback, orgins from the compression of molecular gas clouds along the path of galactic-scale outflows \citep{2017Natur.544..202M, 2019MNRAS.485.3409G}. In Figure \ref{fig4}i, the younger stellar population along the minor axis (red solid profile) of SFOs, compared to the major axis (red dashed line), also proves the birth of young stars in the galactic-scale outflows.

\section{SUMMARY}
\label{sec: summary}

In this paper, we study the physical properties of the star forming galaxies which host galactic-scale outflows (SFOs). The control galaxies (CGs) are matched in the three dimensional parameter space of stellar mass, star formation rate and inclination angle.

$\bullet$ We firstly compare the global properties between the SFOs and CGs. The SFOs are smaller in size and host older stellar population. However, the SFR surface density in the center of SFOs is higher than the CGs. Combining the asymmetry of gas kinematics, these SFOs could undergo gas accretion or galaxy interaction recently which triggers star formation in the center.

$\bullet$ The difference in the stellar population properties along axes between SFOs and CGs describes more detailed star formation progress. The negative slope of $\Sigma_{\star}$ and metallicity gradients in SFOs and CGs, as well as the $Dn4000$ gradient in CGs support the inside-out formation scenario. While the positive $Dn4000$ gradient and higher $\Sigma_{\rm SFR}$ in SFOs imply the recent star formation in the galactic center. Combining the kinematic property gradients of stars and gas, we suggest that the accreted gas triggers star formation in the center of SFOs and further drives galactic-scale outflows.

$\bullet$ By comparing the stellar population properties in galactic-scale outflows (minor axis) and along stellar disk (major axis), we find enhanced star formation and metallicity in the galactic-scale outflows. Those are robust evidences for the positive feedback and metal entrainment in the galactic-scale outflows.
\\ \hspace*{\fill} \\
{\noindent \bf Acknowledgements}
This work is supported by the National Key R\&D Program of China (No. 2017YFA0402700), the National Natural Science Foundation of China (NSFC grants 11733002, 11922302, 11873032, U2038103) and the Research Fund for the Doctoral Program of Higher Education of China (No. 20133207110006). DB is partly supported by RSCF grant 19-12-00145.

Funding for the Sloan Digital Sky Survey IV has been provided by the Alfred P. Sloan Foundation, the U.S. Department of Energy Office of Science, and the Participating Institutions. SDSS- IV acknowledges support and resources from the Center for High-Performance Computing at the University of Utah. The SDSS web site is www.sdss.org. SDSS-IV is managed by the Astrophysical Research Consortium for the Participating Institutions of the SDSS Collaboration including the Brazilian Participation Group, the Carnegie Institution for Science, Carnegie Mellon University, the Chilean Participation Group, the French Participation Group, Harvard-Smithsonian Center for Astrophysics, Instituto de Astrof\'{i}sica de Canarias, The Johns Hopkins University, Kavli Institute for the Physics and Mathematics of the Universe (IPMU) / University of Tokyo, Lawrence Berkeley National Laboratory, Leibniz Institut  f\"{u}r Astrophysik Potsdam (AIP), Max-Planck-Institut  f\"{u}r   Astronomie  (MPIA  Heidelberg), Max-Planck-Institut   f\"{u}r   Astrophysik  (MPA   Garching), Max-Planck-Institut f\"{u}r Extraterrestrische Physik (MPE), National Astronomical Observatory of China, New Mexico State University, New York University, University of Notre Dame, Observat\'{o}rio Nacional / MCTI, The Ohio State University, Pennsylvania State University, Shanghai Astronomical Observatory, United Kingdom Participation Group, Universidad Nacional  Aut\'{o}noma de M\'{e}xico,  University of Arizona, University of Colorado  Boulder, University of Oxford, University of Portsmouth, University of Utah, University of Virginia, University  of Washington,  University of  Wisconsin, Vanderbilt University, and Yale University.
\\ \hspace*{\fill} \\
{\noindent \bf Data Availability}
The data underlying this article will be shared on reasonable request to the corresponding author.

\bsp

\label{lastpage}
\end{document}